\documentclass[conference]{IEEEtran}
\IEEEoverridecommandlockouts
\usepackage{cite}
\usepackage{amsmath,amssymb,amsfonts}
\usepackage{algorithmic}
\usepackage{graphicx}
\usepackage{textcomp}
\usepackage{xcolor}
\usepackage{subcaption}
\usepackage{comment}
\def\BibTeX{{\rm B\kern-.05em{\sc i\kern-.025em b}\kern-.08em
    T\kern-.1667em\lower.7ex\hbox{E}\kern-.125emX}}
\begin{document}

\title{Conflict Analysis and Resolution of Safety and Security Boundary Conditions for Industrial Control Systems \\
\thanks{National Science Foundation Computer and Information Science and Engineering (CISE), award number 1846493 of the Secure and Trustworthy Cyberspace (SaTC) program: Formal TOols foR SafEty aNd. Security of Industrial Control Systems (FORENSICS)}
}

\author{\IEEEauthorblockN{1\textsuperscript{st} Chidi Agbo}
\IEEEauthorblockA{\textit{Computer Science Department} \\
\textit{Boise State University}\\
Boise, United States \\
chidiagbo@u.boisestate.edu}
\and
 \IEEEauthorblockN{2\textsuperscript{nd} Hoda Mehrpouyan}
 \IEEEauthorblockA{\textit{Computer Science Department} \\
 \textit{Boise State University}\\
 Boise, United States \\
hodamehrpouyan@boisestate.edu}
}

\maketitle

\begin{abstract}
 Safety and security are the two most important properties of industrial control systems (ICS), and their integration is necessary to ensure that safety goals do not undermine security goals and vice versa. Sometimes, safety and security co-engineering leads to conflicting requirements or violations capable of impacting the normal behavior of the system. Identification, analysis, and resolution of conflicts arising from safety and security co-engineering is a major challenge, an under-researched area in safety-critical systems(ICS). This paper presents an STPA-SafeSec-CDCL approach that addresses the challenge. Our proposed methodology combines the STPA-SafeSec approach for safety and security analysis and the Conflict-Driven Clause
Learning (CDCL) approach for the identification, analysis, and resolution of conflicts where conflicting constraints are encoded in satisfiability (SAT) problems. We apply our framework to the Tennessee Eastman Plant process model, a chemical process model developed specifically for the study of industrial control processes, to demonstrate how to use the proposed method. Our methodology goes beyond the requirement analysis phase and can be applied to the early stages of system design and development to increase system reliability, robustness, and resilience. 
\end{abstract}


\section{Introduction}
The advancement of technology and the introduction of industry 4.0 have increased complexity in the design and development of cyber-physical systems (CPS). CPS combines both hardware and software resources for computational, communication, and control purposes that are co-designed with the physically engineered components. Examples of CPS include industrial control systems (ICS) that monitor and control physical processes with feedback loops, where physical processes affect computation and vice versa \cite{lee2015past}. The deep integration of discrete computing and continuous physical processes poses major engineering challenges, including the co-engineering of safety and security requirements.  Safety and security are two important properties of an ICS. During the early stages of the design and development of a CPS, safety and security were treated as separate entities. During 2014 and 2015, Chockalingam \cite{chockalingam2016integrated} argued that the research community began to realize the importance of integrating safety and security into CPS. Consequently, research funding and educational projects that focus on safety and security in CPS amount to approximately $\$34, 000, 000$ each year in the US~\cite{nsf2016}.

Safety engineering studies the potential accidents\footnote{Accidents are unintentional or unplanned events that cause an adverse effect} of a system that could lead to hazards\footnote{Hazard is any condition capable of causing human injury or damage to equipment or to the environment} and losses. In ICS, losses are classified as acceptable and unacceptable losses. Safety approaches focus on measures to prevent unacceptable losses, such as equipment damage, loss of life, or financial loss. On the contrary, security engineering focuses on mitigating risks\footnote{Risk is the likelihood that something bad might occur}, threats\footnote{Threat is a potential risk or vulnerability that can impact the system negatively when exploited by a malicious actor}, and vulnerabilities\footnote{Vulnerability is any weakness that poses threat to the system} that exist within the target system. CPS has been classified into safety-critical systems and security-critical systems. In a safety-critical system, safety takes precedence over security and vice  versa,  with different modeling and analysis tools.

Integrating safety and security into ICS is needed to protect the system safely and securely against potential accidents and attacks. However, the combination of safety and security sometimes introduces a conflict, a challenge that has not been fully addressed by the scientific community or industry. kavallieratos et al.\cite{kavallieratos2020cybersecurity} in their comprehensive study of safety and security co-engineering have pointed out that the areas of identifying and resolving conflicting safety and security requirements are under-researched. Conflicting requirements, if not resolved, place an ICS in vulnerable states that can be exploited by cybercriminals.

In 1987, Leveson \cite{leveson1987safety} adopted the use of Petri net modeling over time for the safety analysis of a critical CPS. Other research works that separate safety from security include (\cite{leveson2003applying, burmester2012modeling, zhao2021safety, wang2019avionics, dakwat2018system, wu2016survey, 10.1145/3176258.3176949}). The concept of integration of safety and security analysis gained deep recognition in 2015 with a focus on unification and integration approaches where safety and security are handled as different units and later merged during the life cycle of system development\cite{chockalingam2016integrated, 9847508,9704886}. In practice, we observed that safety and security co-engineering should be part of the early stages of system design and development.

This paper is directed toward safety and security analysis of ICS, focusing on safety and security co-engineering, identification, and resolution of conflicting constraints that expose the system to failure or attacks. The overall goal is to increase the reliability, robustness, and resiliency of an ICS. To this end, we propose the integration of the STPA-SafeSec approach for safety and security analysis and the Conflict-Driven Clause Learning (CDCL) approach for the identification, analysis, and resolution of conflicts. The reasons for our approach are (1). STPA-SafeSec is a systematic technique that adopts a top-down approach to perform an integrated safety and security analysis that captures abnormal behaviors of the system due to component interaction failure. (2). The CDCL approach has been shown to be the best modern SAT technique for the detection, analysis, and resolution of conflicts. 

Our work focuses mainly on the analysis and resolution of conflicts that arise from bounded constraints in safety and security and does not cover all aspects of conflicts that exist in the integration of safety and security, such as confidentiality and/or availability-related conflicts, with a Python script for our implementation. The rest of this paper is organized as follows. Section II discusses related work. Our proposed methodology is presented in Section III. We bring our methodology to the Tennessee Eastman Plant Process in Section IV. Conclusions and future work are discussed in Section V.

\begin{figure*}[h!]
     \centerline{\includegraphics[scale=0.55]{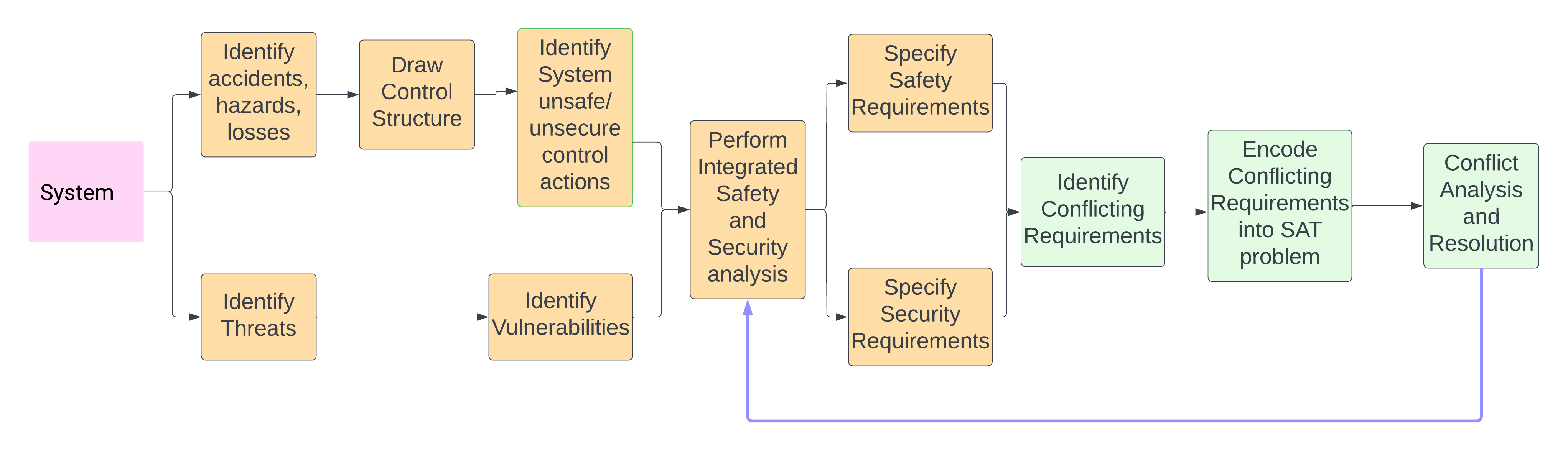}}
    \caption{Proposed Methodology}
    \label{fig-model2}
\end{figure*}
\section{Related Work}
Kornecki et al. \cite{kornecki2010safety} provided a mutual relationship between safety and security in industrial control systems. Their work focused on protecting the software system and the operating environment by proposing the adoption of a safety shell in building security programs. Bruner et al. \cite{brunner2017towards} proposed an integrated model for CPS safety and security requirements based on requirement engineering, risk management, and evidence documentation to support the re-certification process. Bamberger and Martin \cite{bramberger2020co} proposed a process for integrating safety and security according to ISO 26262 standards. ISO 26262 framework provides guidelines for developing safe automotive applications. Cimatti et al. \cite{cimatti2014combining} work goes beyond safety and security integration to develop an approach to verifying safety and security requirements to detect system dependencies and failures.

The analysis and management of the safety risks were performed independently from the security counterpart. However, as the interconnectivity and complexity of CPS increased, the research community recognized the need to combine the analysis of safety and security-related risks. Hayakawa et al. \cite{hayakawa2018proposal} proposed a risk analysis method that deals with safety and security based on the STPA approach. They applied this technique to a medical device (insulin pump) for diabetic patients to uncover accidents that cannot be prevented by functional safety. Kriaa et al. \cite{kriaa2015model} present an S-cube approach to joint risk assessment that is useful for different stages of system development. Reichenbach et al. \cite{reichenbach2012pragmatic} proposed an approach that extended threat vulnerability and risk assessment (TVRA) for integrated security and safety risk analysis. Gleirscher et al. \cite{gleirscher2022challenges} identified technical and sociotechnical challenges in the coassurance of safety and security for collaborative industrial robots (Cobots) without specifying approaches to identify, analyze and resolve risks / conflicts related to safety and security. For example, no testing approach was proposed for Cobots to minimize their overall safety and security risks.

Furthermore, the majority of research work done in this area focused mainly on safety and security co-engineering in ICS with very little work on how to identify and resolve conflicts that arise as a result of their integration. Menon and Vidalis \cite{menon2021towards} present various CPS domains where safety and security conflicts can exist in the system domains, but that was just a pointer.  They suggested an in-depth security approach for conflict resolution that relies on providing multiple layers or overlapping safety and security mitigations, but this method can introduce redundancy and/or more conflicts in the system. Gu et al. \cite{gu2015extracting} proposed an approach for the analysis of safety and security requirements and conflict resolution in industrial control systems. They used risk-based scoring provided by domain experts for conflict resolution, which can be misleading, as the impact of an action or event can sometimes be underestimated, leading to low scoring. Furthermore, this approach cannot be applied to unrelated safety and security requirements or design and implementation phases. Farooq et al. \cite{al2019iotc} present a formal approach to identify conflicting actions between the controller and actuators in an IoT environment. In Sun et al.\cite{sun2009addressing}, they proposed a methodology to find conflicting requirements in automatic door systems. Other related research focused on the automotive and aviation industries with little focus on ICS. They include (\cite{li2022cooperative}, \cite{kornienko2016methodological} , \cite{yang2016static}, \cite{lu2013conflict}). Our work is different in that it goes beyond the requirement analysis phase to system design and implementation using a case study that reflects a real-world ICS. In addition, we provide a systematic conflict analysis and resolution approach that uses the concept of a conflict-driven clause learning approach.

\section{Methodology}
In this work, we propose a methodology based on the STPA-SafeSec-CDCL approach that integrates the STPA-SafeSec~\cite{young2013systems} and conflict-driven clause
learning (CDCL)\cite{marques2021conflict} approaches for the analysis of safety and security, the detection of conflicts, and resolution. This approach applies to the analysis of safety and security in the requirement phase, the design of the system, or the implementation phase. 

Figure \ref{fig-model2} represents our proposed methodology applied to our case study in the next section.
\subsection{The STPA-SafeSec-CDCL Approach}
The STPA-SafeSec technique was developed on top of the STPA \cite{thomas2013systems} and STPA-Sec approaches to address the deficiencies of STPA and STPA-Sec that perform safety and security analysis separately. STPA-SafeSec provides an approach to integrated safety and security analysis from a top-down perspective. Detects both component failures and component interaction failures. Identifies security vulnerabilities\footnote{Security vulnerabilities are deficiencies, errors or faults found within the system security} and requirements\footnote{Security requirements are protection capabilities that need to be satisfied in order to achieve the security of a system} such as scenarios that lead to the violation of security and safety constraints, and uses the results to refine the concept of the system\footnote{System concept denotes the integration and interdependencies of system components and their interactions.} to be safer and more secure\cite{young2013systems}. Furthermore, STPA-SafeSec maps causal scenarios or vulnerabilities to losses and ensures the development of appropriate controls (constraints) over the behavior of the system \cite{howard2017formal}. On the other hand, conflict-driven clause learning (CDCL) is an algorithm for solving Boolean satisfiable (SAT) problems with great capabilities in conflict analysis and resolution. Our proposed methodology combines the concept of STPA-SafeSec and the CDCL approach for co-engineering of safety and security, conflict detection, and resolution. One of the major advantages of our proposed methodology over other methodologies is the fact that it takes into consideration the conflicts or identification of conflicts that may arise from safety and security co-engineering and/or from system design and implementation phases.  

The STPA-SafeSec-CDCL methodology consists of six phases. The first three phases focus on a detailed analysis of safety, security and their integration. Other phases consist of identifying, analyzing, and resolving conflicts that exist during integration, thus specifying mitigation strategies. In this section, we will discuss in detail each of the six phases of our proposed framework.
\subsubsection{Perform In-depth Safety Analysis}
This is the first phase of the STPA-SafeSec-CDCL approach. In this phase, the control structure of the system is analyzed in order to identify and specify possible control actions and/or unsafe control actions. This phase enables domain experts to identify and classify accidents or hazards that lead to acceptable or unacceptable losses, as presented in Table~\ref{table1}. Phase 1 allows for the mapping of control actions to the component layer or the process variables they control. This phase is described in detail in \cite{leveson2018stpa}.
\subsubsection{Perform In-depth Security Analysis}
This phase focuses on the identification and classification of various threats and vulnerabilities to the system.   In-depth security analysis tends to analyze and uncover threats related to the confidentiality, integrity, and availability of the system. In addition, this phase equips security experts with the necessary knowledge about the system to find weaknesses that can threaten the security of the system. Examples of such threats include command injection, command manipulation, command delay or drop, measurement drop or manipulation, etc. as explained in section \ref{case_study}.
\subsubsection{Perform Integrated Safety and Security Analysis}
Here, safety and security experts meet to ensure that security goals improve safety and do not undermine safety goals, and vice versa. Phase 3 is crucial to ensure a safe and secure ICS. Enhances the harmonization of safety and security goals toward the overall safety and security of the system.  safety and security requirements are specified based on accidents, hazards, losses, causal factors, unsafe control actions, threats, and vulnerabilities identified in Phases 1 and 2.

\subsubsection{Identification of Conflicting requirements}
The integration of safety and security analysis can introduce conflicts that are known or unknown to security experts. This phase can be the most difficult, depending on the system being analyzed. Sometimes there are conflicts within the safety and security domain or within their integration. However, design or implementation goals can conflict with safety or security goals. For example, consider system A designed to operate properly under noise with no significant impact on the system. However, the introduction of noise at a certain state of the system can cause an increase or decrease in some process variables in A that conflict with the safety or security constraints or the normal behavior of system A. During the conflict detection phase, all possible scenarios or causes that could lead to a conflict are analyzed to ensure that the system operates in a conflict-free manner. In this paper, we adopt a conflict identification approach proposed by Thomas \cite{thomas2013extending} that is based on four principles (i) source controller issuing control actions (SC), (ii) type of control action (T) such as provided or not provided, (iii) control action (CA), and (iv) context in which control action is provided or not provided (Co). A conflict occurs when providing or not providing a control action leads to hazards/threats. For example, consider an automatic door system(ADS) used in banks, airports, and other critical facilities designed to detect metal objects(e.g., guns) and automatically deny the person an entrance into the building. The security requirement for such a system is designed to ensure that the ADS shuts the door against anyone with metal objects. However, in the event of emergencies, such as fire outbreaks, the ADS will enforce safety requirements for evacuation purposes, thereby compromising security goals. In this case, enforcing or not enforcing safety/security requirements leads to hazards/threats. We add this conflict identification method to our case study in section \ref{case_study}.

\subsubsection{Conflict Analysis and Resolution}
Conflict analysis and resolution are performed when there are conflicting requirements that occur during integration of safety and security or when design goals conflict with safety/security goals. This phase enables domain experts to identify causal factors and conflict resolution strategies. The knowledge gained is needed to redefine the safety and security constraints or the goals of system design. Our methodology adopts the CDCL approach to conflict analysis and resolution, discussed in detail in section \ref{CDCL}

\subsubsection{ Redefine Safety and Security Constraints, and/or Mitigation strategies}
In this phase, safety and security requirements specified in phase 3 are re-evaluated to meet the system's current state based on the knowledge gained from the conflict analysis and resolution phase. The safety and security constraints are redefined to ensure conflict-free constraints. Mitigation strategies can also be specified to address the overall safety and security goals of the system.

\subsection{STPA-SafeSec Approach}
The STPA-SafeSec approach is based on the fundamental principles of STPA with an extension to the security domain. The STPA-SafeSec approach addresses the weaknesses of the STPA and STPA-sec frameworks by providing interdependencies between security and safety constraints. The result of the STPA-SafeSec approach helps domain experts identify potential hazards (safety accidents) or threats(security vulnerability) that can cause system loss\cite{friedberg2017stpa}. Although STPA-sec shows that STPA can also be used for security analysis to ensure system safety\cite{young2013systems},\cite{young2014integrated}. However, the STPA approach prioritizes safety over security. On the contrary, STPA-SafeSec provides a unified and integrated approach that gives equal relevance to both the safety and security of the system and the reason for its use in this paper. In STPA-SafeSec, safety and security constraints are specified based on the control structure of the system. Furthermore, in the STPA-SafeSec analysis, we first identify system losses and hazards, control structures, threats, and / or vulnerabilities, and safety and security constraints. We applied this approach to our case study in section \ref{case_study}.

\subsection{Conflict Driven Clause Learning Approach (CDCL) \label{CDCL}}
The CDCL method is similar to the DPLL approach with the standard backtracking search procedure where unit propagation is executed after each decision assignment, creating a new decision level \cite{marques2021conflict}. The CDCL approach follows a non-chronological backtracking search technique in which backtracking occurs once a conflict is identified and, during conflict resolution, new clauses are learned. The CDCL approach jump starts at the appropriate decision level during backtracking when conflicts occur. The CDCL approach operates on clauses in the form of a conjunctive normal form(CNF). Clauses are literals. A literal is a variable a or its negation $\neg{a}$. A CNF represents a conjunction of disjunctive clauses\footnote{Disjunctive clauses are clauses consisting of OR-ed literals}. For example, consider a set of literals a, b, $\neg{a}$ that takes a true (1) or false(0) value. Clauses (a $\vee$ b) and ($\neg{a}$ $\vee$ b) are of disjunctive form. Therefore, a CNF denotes clauses of the form (a $\vee$ b) $\wedge$ ($\neg{a}$ $\vee$ b). 
\textbf{Definition-1 Conflict:} Consider a CNF formula $\psi =$ (a $\vee$ b) $\wedge$ ($\neg{a}$ $\vee$ b) such that an assignment of true value to variable a and false to variable b leads to an unsatisfiable result (UNSAT). A conflict occurs when a variable a or b is assigned true and false values in a given formula.
A CDCL solver returns the UNSAT result if and only if it is unable to resolve the conflict by finding an assignment that satisfies the formula and otherwise returns a SAT result. During the assignment of variables, CDCL adopts the concept of unit propagation or Boolean constraint propagation(BCP) to determine the Boolean value that must be assigned to a particular variable for a formula to be satisfied. For example, consider the formula $\alpha = $ (a $\vee$ b $\vee$ c $\vee$ d). If we assign a false value to variables a, b, and c, by unit propagation or BCP, CDCL will assign the true value to variable d for $\alpha$ to be satisfiable. If it is a unit clause, CDCL will assign a true value to the variable for it to be satisfiable.

\begin{figure}[hbt!]
\centering
\begin{subfigure}{0.4\textwidth}
    \centering
    \includegraphics[scale = 0.55]{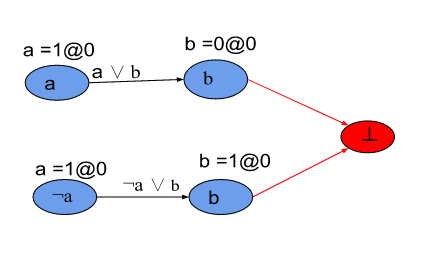}
    \caption{Implication Graph for $\psi$}
    \label{fig:Implication Graph}
\end{subfigure}
\begin{subfigure}{0.4\textwidth}
    \centering
    \includegraphics[scale = 0.55]{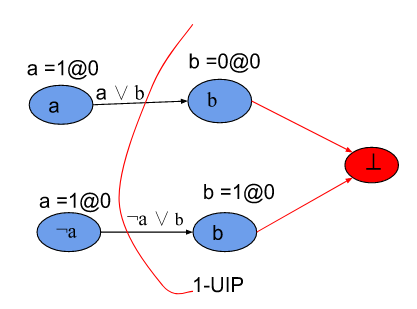}
    \caption{First-UIP for $\psi$}
    \label{fig:First-UIP}
\end{subfigure}
\begin{subfigure}{0.4\textwidth}\quad 
    \centering
    \includegraphics[scale =0.60]{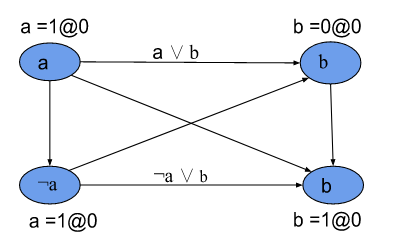}
    \caption{Resolution Graph for $\psi$ }
    \label{fig:Resolution Graph}
\end{subfigure}
\caption{Conflict Analysis and Resolution for formula $\psi$}
\label{Fig-model101}
\end{figure}

Furthermore, in the case of a conflict, CDCL backtracks to the appropriate decision level made before the conflict occurred. During conflict analysis and resolution, implication graphs are created to help analyze and visualize the assignment of variables or decisions that caused a conflict. It is a powerful graph for conflict analysis and resolution, as we can easily reverse the last decision made before a conflict, undo the assignment, and reassign variables again until the conflict is resolved. During conflict resolution, new clauses are learned. The modern CDCL SAT solver implements the concept of the first unique implication point (UIP) for clause learning. UIP represents the node or path closest to the conflicting node in the implication graph. The use of first UIPs is necessary to reduce the learning of new redundant clauses.  Let C\textsubscript{1} and C\textsubscript{2} represent the clauses (a $\vee$ b)  and ($\neg{a}$ $\vee$ b) in $\psi$, respectively. Then, the implication graph for the CNF formula $\psi$ is shown in Figure~\ref{fig:Implication Graph}.

In the implication graph, as shown in Figure~\ref{fig:Implication Graph}, we assign true(1) and false(0) values to variables \textbf{a} and \textbf{b}, respectively, at a decision level 0 for C\textsubscript{1} to be satisfiable. If we maintain the same assignment for C\textsubscript{2}, C\textsubscript{2} becomes unsatisfiable, thus causing the general formula $\psi$ to return the UNSAT result. Therefore, to satisfy $\psi$, we need to assign a true (1) value to the variable \textbf{b} that contradicts the initial assignment of $\textbf{b} = false(0)$ leading to a conflict. Once a conflict occurs, CDCL jumps back to the decision level that led to the conflict. In our case, we go back to decision level 0 and undo the assignment of \textbf{a}. During the backtracking, new clauses are learned. The 1-UIP as shown in Figure~\ref{fig:First-UIP} helps us to make the right cut for the nodes closest to the conflicting nodes. In our example, the learned clause C is (a $\vee$ $\neg{a}$).  The new clause is called a conflict clause. The conflict clause is added to formula $\psi$ to avoid repeating the same assignment that leads to a conflict. During conflict resolution (Figure~\ref{fig:Resolution Graph}), we reassigned the false value (0) to \textbf{a} and the true value (1) to \textbf{b} to make $\psi$ satisfiable. 

\begin{figure}[hbt!]
     \centerline{\includegraphics[scale=0.4]{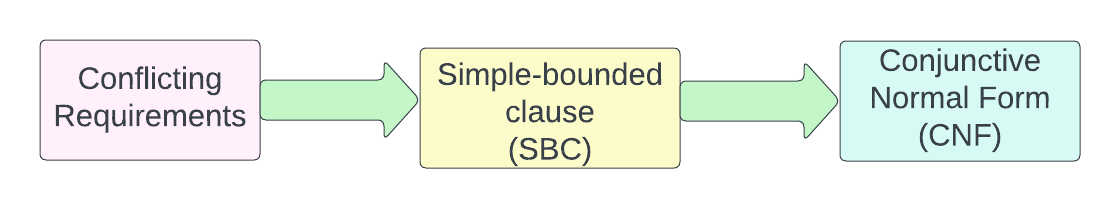}}
    \caption{Encoding Conflicting requirements into SAT problem}
    \label{fig-model12}
\end{figure}

Safety and security requirements in ICS are often represented as bounded constraints, where the system and its process variables are expected to be within a set of intervals and not as a Boolean or CNF formula. Therefore, one of the intriguing parts of this section is how to encode conflicting requirements into SAT problems.  Therefore, we propose a method(see Figure \ref{fig-model12}) to encode bounded constraints in a SAT problem.

In this method, the safety or security requirements are first converted to a simple bounded clause (SB) and then to a CNF. The concept of SB has been applied in the areas of linear measurement\cite{foucart2013invitation}, stochastic systems for calculating the fluctuation of time-based currents\cite{garrahan2017simple}, nonlinear control systems for computing a set of reachable states based on differential inequalities \cite{scott2013bounds}, interval-based arithmetic constraints\cite{franzle2006efficient}, machine learning for computing simple bounds for interval-valued covariates in linear regression\cite{schollmeyer2021computing}, artificial intelligence for defining and controlling sub-spaces of a network of agents\cite{zhang2013upper}, etc. Our encoding method is similar to the work done by \cite{scheibler2017applying}. To the best of our knowledge, our work is the first to lift the concept of a simple bound to safety and security constraints in ICS for the identification of conflicts and resolution.  We have implemented encoding of conflicting requirements for bounded constraints in Python\footnote{ https://github.com/Chidi93/IEEE6thConfItaly.git}and it is used in our case study.

\section{Case Study} \label{case_study}
The Tennessee Eastman challenge process (TEP) is the simulation of a real chemical process developed specifically for the study of industrial control processes. We choose the TEP plant for 3 main reasons (i) The TEP plant is a widely used plant process model for the study of CPS\cite{chen1998predictive}, \cite{capaci2019revised}. (ii)It is made up of various components, levels, and process variables found in today's chemical plants such as reactor, compressor, stripper, condenser, separator, analyzers, sensors, actuators(valves), feed components, pressure, temperature, etc. and (iii) TEP has been used in the study of CPS security and attack detection \cite{krotofil2014cps}, \cite{liu2020toward}, \cite{segovia2020cyber}, \cite{gao2021anomaly}, \cite{ghafouri2018adversarial}, etc.
\begin{figure}[hbt!]
     \centerline{\includegraphics[scale=0.35]{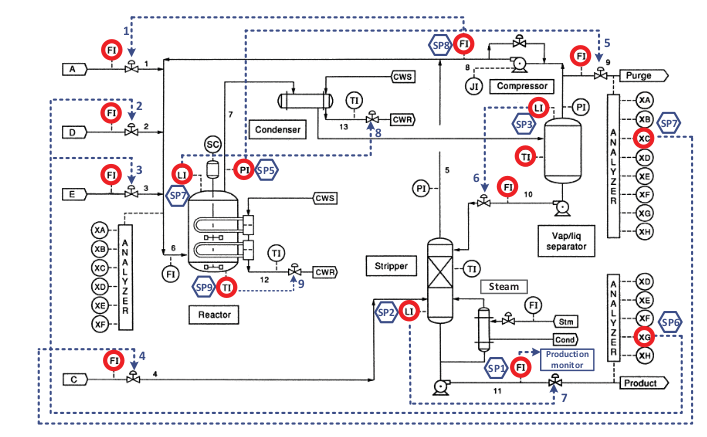}}
    \caption{The Tennessee Eastman Plant \cite{krotofil2014}}
    \label{fig-model4}
\end{figure}

The plant contains eight chemical components comprising four reactants (A, C, D, and E), two liquid products, one by-product, and one inert component as seen below;

$A(\textsubscript{g}) + C(\textsubscript{g}) + D(\textsubscript{g}) \rightarrow G(\textsubscript{liq}) \Rightarrow Product 1 $

$A(\textsubscript{g}) + C(\textsubscript{g}) + E(\textsubscript{g}) \rightarrow H(\textsubscript{liq}) \Rightarrow Product 2 $

In our study, we implemented our proposed methodology -the STPA-SafeSec-CDCL framework - in the TE Chemical reactor control system (CRCS) described in \cite{wang2019avionics}. The TE chemical reactor is the unit of the TE plant where the chemical reactions for the creation of products G and H take place. In the STPA-SafeSec-CDCL approach, the first phase is to perform an in-depth safety analysis of the CRCS system to identify losses and hazards at the system level (see Table \ref{table1}). 

\begingroup
\setlength{\tabcolsep}{10pt} 
\renewcommand{\arraystretch}{1.5} 
\begin{table}[h!]
\begin{tabular}{ |p{3cm}||p{0.7cm}|p{3cm}| }
 \hline
 Losses & Reference & Refers-To \\
 \hline
 Loss of life or injury   & [L-1]    & \\
 Loss of or damage to equipment   & [L-2]    & \\
 Loss of or damage to product   & [L-3]    & \\
 Monetary loss   & [L-4]    & \\
 Contaminated environment & [L-5] & \\
\hline
 Hazards & & \\
 \hline
 Operating reactor beyond set points   & [H-1] & [L-1],[L-2],[L-3],[L-4],[L-5]\\
 Plant inability to maintain process variables within defined threshold during chemical production & [H-2] & [L-1],[L-2],[L-3],[L-4],[L-5]\\
 Release of high volume of chemicals into the reactor & [H-3] & [L-1],[L-2],[L-3],[L-4],[L-5]\\
 Opening reactor's product and discharge valves at the same time & [H-4] & [L-3],[L-4] \\
 Plant releases contaminated materials  & [H-5] & [L-1],[L-5] \\
 \hline
\end{tabular}
\caption{Losses and Hazards associated with the TE Chemical reactor}
\label{table1}
\end{table}

The next step is to examine the control structure to learn the detailed list of commands, manipulated process variables (MPV), and feedback during chemical production. 
\begin{figure}[hbt!]
     \centerline{\includegraphics[scale=0.4]{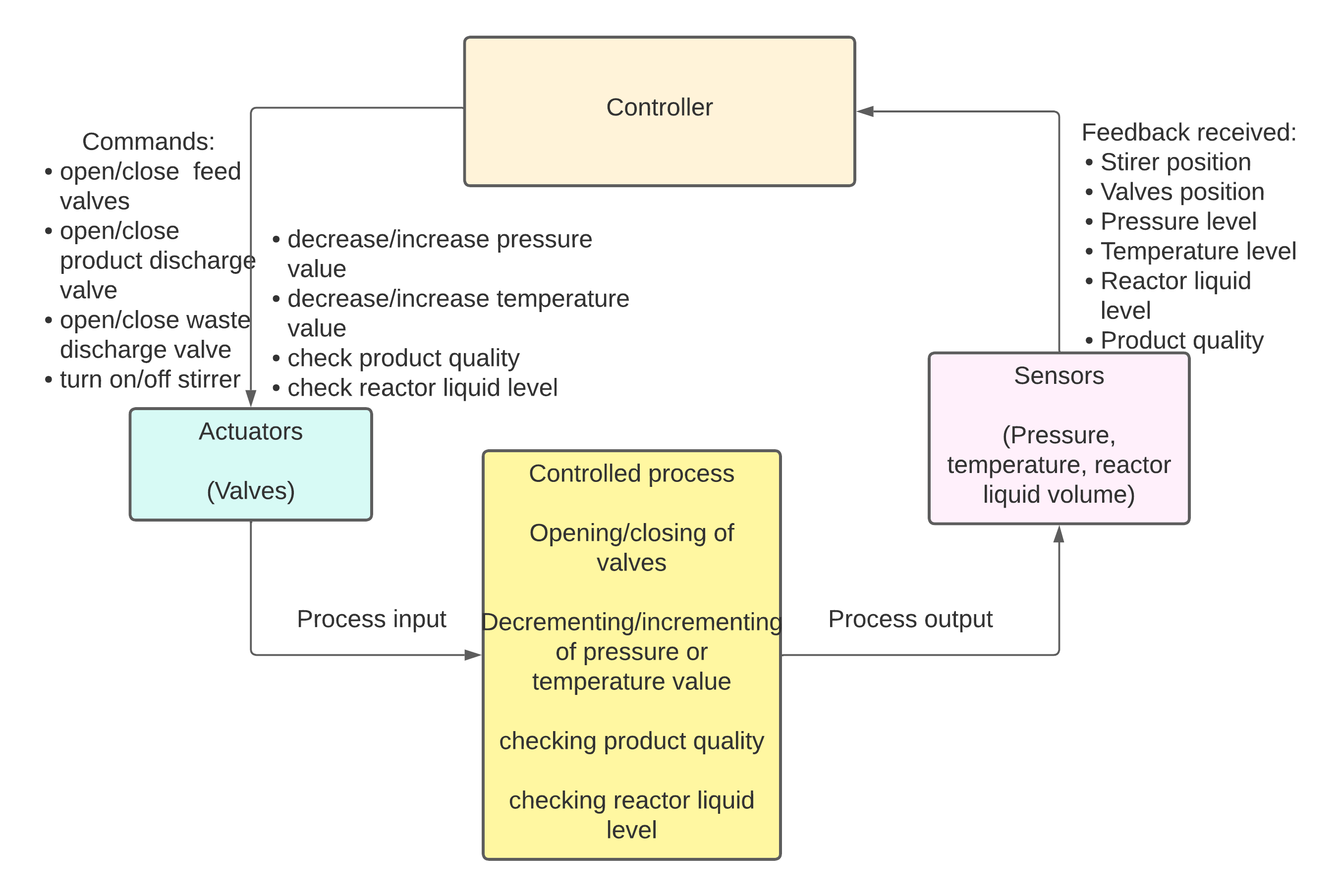}}
    \caption{TE CRCS}
    \label{fig-model5}
\end{figure}

The goals and objectives of the TE control system have been described by downs and Vogel \cite{downs1993plant} as follows: (i) keeping the process variables within the desired values. (ii) ensure that the processes operate within equipment constraints. (iii) Reduce the variations in product rate and process variables during disturbances, (iv) Reduce valve movement that affects other processes, and (iv) recover quickly and smoothly from disturbances, product mix, or production rate changes.

We analyze the CRCS system to identify necessary control actions (CA) and unsafe control actions (UCA) that cause hazards that lead to loss. We applied the four ways of identifying CA and UCA (see Table \ref{table2}) described by Leveson and Thomas \cite{leveson2018stpa}. 
\begingroup
\setlength{\tabcolsep}{10pt} 
\renewcommand{\arraystretch}{1.5} 
\begin{table*}[hbt!]
\begin{tabular}{ |p{3cm}||p{3cm}|p{3cm}|p{3cm}|p{3cm}| }
 \hline
 Control Action & Not providing causes hazard &  Providing causes hazard & Provided too early/late & Stopped too early/late\\
 \hline
 Equipment must be operated within set points (CA-1) &Inability to control reactor operations within set limits leads to shutdown or equipment damage & N/A & Enforcing operating equipment within set points but too late causes damage to equipment or worn out
& Stopping too late when equipment operating out of bound causes damage\\

Maintain process variable within desired values (CA-2) & leads to shutdown or equipment or product damage & N/A & Too late causes equipment shutdown/damage or product damage & N/A\\

Minimize variability in valves (CA-3) & High release of raw materials into the reactor causing reactor overflow or product damage & Opening of product and waste valves at the same time causes reactor overflow or product damage & Closing of the feed or discharge valves too early/late impact the product quality & Opening/closing of the feed valves for too long/early impacts product quality \\
Reduce process variables and feed rates variation under disturbances (CA-4)& causes system shutdown or product damage & N/A & too late leads shutdown or product damage & Stopped too early/late causes shutdown or product damage\\

The plant must recover quickly and smoothly from disturbances and production rate changes (CA-5)& system shutdown, equipment or product damage & N/A & too late causes system shutdown, equipment or product damage & stopped too early / late leads to system shutdown, equipment or product damage\\
 \hline
\end{tabular}
\caption{CRCS Control Actions and their Impact on the System when providing, not providing, providing too early / late or stopping too early or late }
\label{table2}
\end{table*}
\subsection{Safety Requirements}
As shown in Table \ref{table2}, we can derive unsafe control actions and their causes(CS) for each control action.\\
\textbf{CA-1:}Equipment must be operated within set points\\
\textbf{UCA-1:}Equipment is operated beyond set limits.\\
\textbf{CS-1:}Controller fails to regulate the operation of the plant within set points.\\
\vspace{0.1cm}\\
\textbf{CA-2:}Maintain process variables within desired values\\
\textbf{UCA-2:}Operating plant under out-of-bounds process variables\\
\textbf{CS-2.1:}Failure of sensors to properly monitor the current state of the system\\
\textbf{CS-2.2:}Communication gap between sensor and controller.\\
\vspace{0.1cm}\\
\textbf{CA-3:}Minimize frequent movement of the valves\\
\textbf{UCA-3:}Opening/closing of valves without restraint\\
\textbf{CS-3.1:}Controller issue a command that causes the opening of both product and waste valves at the same time due to poor feedback from the sensor.\\
\textbf{CS-3.2:}Sensor's failure to communicate the current state of the valves to the controller.\\
\textbf{CS-3.3:}Actuator's failure to receive commands from the controller.\\
\vspace{0.1cm}\\
\textbf{CA-4:}: Reduce the variation in feed rates under disturbances\\
\textbf{UCA-4:}Activation of disturbances that increase feed rates \textbf{CS-4:} Control system inability to stabilize feed rates under disturbances due to failure or attack.
\textbf{CA-5:} Plant must recover quickly and smoothly from disturbances 
\textbf{UCA-5:} Plant does not recover under disturbances.
\textbf{CS-5:} Increase in the variable of the process under disturbances that cause the plant to shut down.
\subsection{Security Requirements}
Having performed a detailed system safety analysis, we can now perform the security analysis. The security analysis phase is based on security threats(SCT) and vulnerabilities of the target system. In the TE CRCS, we have identified various threats based on the CIA triad-Integrity, Confidentiality, and Availability.
\\
\begin{itemize}
    \item[i)] {\textbf{Confidentiality threats}}
    \item {\textbf{SCT-C-1:} Unauthorized access to plant's proprietary data}\\
    {\textbf{Scenarios/possible attacks:}} An external person or a disgruntled employee has access to the operational data of the plant to exploit the plant.
     \item {\textbf{SCT-C-2:}}Unauthorized access to the HMI or control commands.\\
     {\textbf{Scenarios/possible attacks:}} Remote connection to the HMI or control commands by a malicious actor to manipulate the system.\\
      \item[ii)]{\textbf{Integrity threats}}
        \item {\textbf{SCT-I-1:}}Command Manipulation\\
        {\textbf{Scenarios/possible attacks:}}Remote connection to the controller or HMI to inject or modify control commands.
        \item {\textbf{SCT-I-2:}}Command drop\\
        {\textbf{Scenarios/possible attacks:}}Man in the middle attack (MITM) that drops control commands.
         \item {\textbf{SCT-I-3:}}Measurement Manipulation\\
        {\textbf{Scenarios/possible attacks:}}MITM attack that intercepts and modifies sensor measurement with fake values.
         \item {\textbf{SCT-I-4:}}Measurement drop\\
        {\textbf{Scenarios/possible attacks:}}Man in the middle (MITM) attack that drops sensor measurement.\\
      \item[iii)]{\textbf{Availability threats}}
      \item {\textbf{SCT-A-1:}}Command delay\\
      {\textbf{Scenarios/possible attacks:}} A DOS attack that determines the optimal time to delay control commands and thus drives the plant into an unsafe state.
       \item {\textbf{SCT-A-1:}}Measurement delay\\
      {\textbf{Scenarios/possible attacks:}} A DOS attack that delays sensor measurement so that the controller uses stale data from its memory that do not reflect the current state of the plant to make decisions.
\end{itemize}
\subsection{Safety and Security co-engineering}
Safety and security are two essential properties in ICS, and the importance of their integration cannot be overemphasized. Integration of safety and security analysis is necessary to ensure that safety goals do not undermine security goals and vice versa. When we integrate our results into phases 1 and 2, we can see that unsafe control actions and causes in the Safety domain are synonymous with the identified threats in the Security domain. This phase bridges the gap that exists when the safety analysis is done separately from the security analysis. Safety and security experts collaborate to ensure the overall safe and secure operation of the system. Security experts match threats with possible UCA and their causes to ensure that all potential threats to the system are covered. Security experts re-evaluate the system with safety lenses and vice versa. In this work, we focus on those safety requirements of the TE plant that are represented as bounded constraints. Table \ref{table3} summarizes the safety limits.
\setlength{\tabcolsep}{10pt} 
\renewcommand{\arraystretch}{1.5} 
\begin{table}[hbt!]
\begin{tabular}{ |p{1.2cm}||p{1.5cm}|p{1.2cm}||p{1.5cm}|}
 \hline
 \multicolumn{4}{|c|}{Safety Boundaries/Constraints} \\
 \hline
 Component & Boundary(Feed rates in \%) & Component & Boundary(Feed rates in \%) \\
 \hline
  Feed A   & [24, 30]    &Reactor pressure & [2800, 3000]\\
  Feed C   & [60, 62]    &Stripper level & [46, 54]\\
  Feed D   & [62, 64]    &Reactor Level & [54, 55] \\
  Feed E   & [52, 55]    &Quality & [54, 55] \\
  Product G   & [52, 56]    &Price & [100, 120] \\
  Product H   & [42, 46]    &Production & [22, 23] \\
 \hline
\end{tabular}
\caption{Safety Constraints of the TE Plant }
\label{table3}
\end{table}

Defining safety requirements in ICS as bounded constraints uses a strict boundary approach. The reason for this approach is crucial in the safety and security domain for the rapid detection of out-of-bounds constraints. For example, if A is a component, the bound A$\in[25, 30]$ is tighter than A$\in[0, 30]$ and provides more information on the interval of A. Furthermore, there is a violation or conflict when there are actions or events that cause the system to operate outside the set boundaries. In the case of the TE plant, the control system is designed to ensure that the system operates within the defined set points. For example, the TE plant shuts down when the reactor pressure exceeds 3000 kpa.

While safety helps prevent failure, security protects the system against attacks. Some research work has shown the impact of DOS or integrity attacks on the TE plant and how they successfully attacked the system (\cite{krotofil2014}, \cite{kiss2015denial}, \cite{genge2015system}). On the basis of our analysis, we define the following security requirements to address all identified threats to the system.

\begin{itemize}
 \item[i)] \textbf{Confidentiality}\\ 
\textbf{SC-C-1:}  Only authorized users should access the HMI using strong passwords and multifactor authentication for login. Security tools and monitoring systems must be configured to monitor and prevent attempted logins.\\
 \textbf{SC-C-2:} Plant proprietary data must be encrypted to prevent users from reading the plant’s operational commands and workabilities.
 \item[ii)]\textbf{Integrity} \\ 
\textbf{SC-I-1:}  Every input from the HMI or controller must be validated\\
\textbf{SC-I-2:}  The system should establish an average time required for the controller to communicate with sensors and actuators and vice versa.
\textbf{SC-I-3:} Build a defensive mechanism to prevent attackers from modifying control commands, sensor, actuator values or feed rates.
\textbf{SC-I-4:} Every user must be authenticated before accessing plant facilities or systems.
\item[iii)]\textbf{Availability}\\
\textbf{SC-A-1:} Monitor the traffic of the TE network to identify unusual traffic patterns or unusual communication delays between the controller and sensors / actuators \\
\textbf{SC-A-2:} Make the TE network resistant to DOS attacks by establishing firewalls and placing servers in different data centers.
\end{itemize}
Integration of safety and security analysis can result in conflicting requirements. Based on our adopted conflict identification approach, in our case study, we found that the activation of some disturbances conflicts with the safety/security and typical behavior of the system, as detailed in the next section. From our analysis, we observed that providing or not providing control action 4(CA-4) leads to hazards/threats under disturbance activation because the controller could not maintain the process variables and feed rates variations within the acceptable threshold, thus conflicting with the plant's safety and security goals.
\subsection{Conflicts Analysis and Resolution:}
The introduction of disturbances in chemical plants, such as the TE process model, exposes the plant to noise and variations that could negatively impact the system. The TE plant is designed to operate under disturbances without noticeable effects on the processes, process variables and / or the plant based on the control objectives.  Sometimes designers build systems that can withstand noise or uncertainty as a way to prove the reliability, robustness, and resiliency of the system. However, such a design and implementation could lead to redundancy and violations(conflicts) of system properties.
In the TE plant, we closely observed the operation of the plant under disturbances and found some redundancy and violations in the implementation.

From our analysis, we found that activation of some disturbances violates the safety and normal behavior of the TE plant. To demonstrate the impact of disturbances on the system, we used the TE plant code developed by Bathelt and Ricker \cite{bathelt2015revision}. We run the code in Matlab. In the simulation setup, we set the time T for the simulation at 50 h. First, we run the simulation without disturbance activation using the base case values specified by Downs and Vogel \cite{downs1993plant} and observe the normal behavior of the plant (see Table \ref{table3}). Using the same setup, we performed other simulations with activation of disturbances IDV(1), IDV(11), and IDV(13) and recorded plant behavior.  IDV(1) is a step-type disturbance that manipulates the feed ratios A and C.  IDV(11) is a random variation disturbance that affects the temperature of the reactor cooling water inlet, and IDV(13) is a slow drift-type disturbance that affects reaction kinetics. The specified control actions ensure that the Plant operates properly under disturbances.

Unfortunately, from the simulation result, we observed that disturbance activation conflicts with the safety and security goals of the plant, causing the plant to exhibit abnormal behavior. For example, the production cost increased more than 150\% higher than the normal production cost. There was a significant decrease in the production rate and a great increase in the feed rates. The reactor pressure and stripper recorded high and low peak levels, respectively, leading to the shutdown of the plant. Table \ref{table4} summarizes the impact of disturbance activation on the system. It is important to note that such fluctuations can have catastrophic consequences in chemical, nuclear or water treatment plants.

\setlength{\tabcolsep}{10pt} 
\renewcommand{\arraystretch}{1.5} 
\begin{table}[hbt!]
\begin{tabular}{ |p{1.2cm}||p{1.5cm}|p{1.2cm}||p{1.5cm}|}
 \hline
 \multicolumn{4}{|c|}{Chemical production under disturbances IDV(1)} \\
 \hline
 Component & Boundary(Feed rates in \%) & Component & Boundary(Feed rates in \%) \\
 \hline
  Feed A   & [28, 100]    &Reactor pressure & [2760, 2820]\\
  Feed C   & [55, 61]    &Stripper level & [30, 70]\\
  Feed D   & [62, 64]    &Reactor Level & [62, 68] \\
  Feed E   & [52, 55]    &Quality & [54, 55] \\
  Product G   & [52, 56]    &Price & [50, 250] \\
  Product H   & [42, 46]    &Production & [22, 23] \\
 \hline
  \multicolumn{4}{|c|}{Chemical production under disturbances IDV(11)} \\
 \hline
  Feed A   & [28, 100]    &Reactor pressure & [2780, 2960]\\
  Feed C   & [57, 61]    &Stripper level & [-30, 50]\\
  Feed D   & [63, 64]    &Reactor Level & [64, 69] \\
  Feed E   & [53, 60]    &Quality & [54, 58] \\
  Product G   & [54, 58]    &Price & [50, 300] \\
  Product H   & [37, 44]    &Production & [20, 23] \\
 \hline
 \multicolumn{4}{|c|}{Chemical production under disturbances IDV(13)} \\
 \hline
  Feed A   & [10, 45]    &Reactor pressure & [2500, 2900]\\
  Feed C   & [60, 63]    &Stripper level & [10, 80]\\
  Feed D   & [63, 64]    &Reactor Level & [64, 69] \\
  Feed E   & [52, 56]    &Quality & [50, 57] \\
  Product G   & [54, 58]    &Price & [40, 300] \\
  Product H   & [37, 44]    &Production & [20, 22] \\
 \hline
\end{tabular}
\caption{Plant production under disturbance activation}
\label{table4}
\end{table}

Having identified conflicting security and safety boundaries in our system, as presented in Table~\ref{table4}, we applied the encoding method described in our methodology. We first encoded the constraints into a simple bound clause and then translated the SB-clause into CNF solvable by the CDCL approach. For example, let D\textsubscript{0} be a domain that represents the overall chemical reactions in the TE reactor that produce the products G and H with or without disturbances. Let D\textsubscript{1} be a subdomain that represents the chemical reactions in the reactor that produce the product G under normal plant operation (without disturbances) such that \newline
D\textsubscript{1} = $A\textsubscript{(g)} + C\textsubscript{(g)} + D\textsubscript{(g)} = G\textsubscript{(liq)}$ \hspace{0.5cm} (i)

Let D\textsubscript{2} be a subdomain that represents the chemical reactions in the reactor that produce the product H under normal plant operation (without disturbances) such that: \newline
D\textsubscript{2} = $A\textsubscript{(g)} + C\textsubscript{(g)} + E\textsubscript{(g)} = H\textsubscript{(liq)}$ \hspace{0.5cm} (ii) \\
Let $D\textsubscript{3}$ and $D\textsubscript{4}$ be two subdomains that represent the chemical reactions in the reactor that produce products G and H, respectively, under disturbances such that: \\
D\textsubscript{3} = $\neg{A\textsubscript{(g)}} + \neg{C\textsubscript{(g)}} + \neg{D\textsubscript{(g)}} = \neg{G\textsubscript{(g)}}$ \hspace{0.2cm} (iii) \newline
D\textsubscript{4} = $\neg{A\textsubscript{(g)}} + \neg{C\textsubscript{(g)}} + \neg{E\textsubscript{(g)}} = \neg{H\textsubscript{(g)}}$ \hspace{0.2cm} (iv) \newline
Note that the negation of a component denotes that the lower limit or upper limit of the component or both limits do not fall within acceptable safety limits or threshold.

Now, let us decompose the components that make up D\textsubscript{1}, D\textsubscript{2}, D\textsubscript{3}, and D\textsubscript{4} with their respective limits. For D\textsubscript{1}, we have that: \newline
A$\in[24, 30]$, C$\in[60, 62]$, D$\in[62, 64]$

Let l1, l3, l5, and l7 represent the lower bounded safety properties, and l2, l4, l6, and l8 represent the upper bounded safety properties such that:

$D\textsubscript{1} = ((a\ge{24}) \wedge (a\le{30})) \vee ((c\ge{60})\\ \wedge (c\le{62})) \vee ((d\ge{62}) \wedge (d\le{64}))$ and\\
$l\textsubscript{1}\leftrightarrow{((a\ge{24})}, l\textsubscript{2}\leftrightarrow{((a\le{30})}, l\textsubscript{3}\leftrightarrow{((c\ge{60})},l\textsubscript{4}\leftrightarrow{((c\le{62}), l\textsubscript{5}\leftrightarrow{((d\ge{62})}, l\textsubscript{6}\leftrightarrow{((d\le{64})}}$

So, we have: \newline
$D\textsubscript{1} = (l1 \wedge l2) \vee (l3 \wedge l4) \vee (l5 \wedge l6)$  \hspace{1.7cm} (1)\\
$D\textsubscript{2} = (l1 \wedge l2) \vee (l3 \wedge l4) \vee (l7 \wedge l8) $ \hspace{1.7cm} (2)\\
$D\textsubscript{3} = (l1 \wedge \neg{l2}) \vee (\neg{l3}\wedge l4) \vee (l5 \wedge l6)$ \hspace{1.2cm} (3)\\
$D\textsubscript{4} = (l1 \wedge \neg{l2}) \vee (\neg{l3} \wedge l4) \vee (l7 \wedge \neg{l8})$ \hspace{1cm} (4)\\

Therefore, $D\textsubscript{0} = (D\textsubscript{1} \vee D\textsubscript{2}) \wedge (D\textsubscript{3}  \vee D\textsubscript{4}) \equiv (D\textsubscript{1} \wedge D\textsubscript{3} ) \vee (D\textsubscript{2} \wedge D\textsubscript{4})$

We pass D\textsubscript{0} to our Python script\footnote{see details here: https://github.com/Chidi93/IEEE6thConfItaly.git} that converts it to a CNF formula. The script converts the individual domains to CNF. It also contains methods for creating implication graphs. The script takes a formula with variables A, B, C, ..., H which represents l1, l2, l3, ..., l8 (in our case) and converts it to a CNF formula. 

During conflict analysis, we use an implication graph. In our analysis, we created implication graphs for IDV (1), IDV (11) and IDV (13). It is imperative to note that the assignment of values to variables l1, l2, ..., l8 is such that a true value denotes a lower bound or an upper bound within the safety boundary, otherwise false. We adopted this assignment method on the basis of knowledge of the underlying system. 

\begin{figure}[hbt!]
  \begin{subfigure}{0.5\textwidth}
    \centering
    \includegraphics[width=\linewidth]{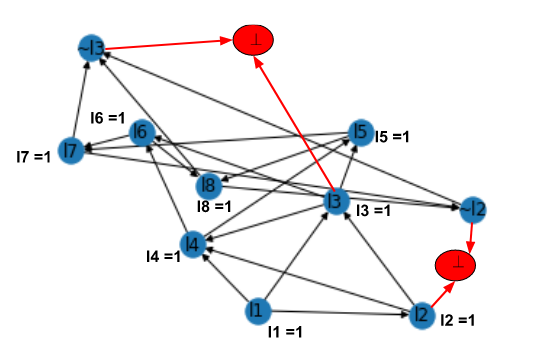}
    \caption{ \scriptsize Implication graph for IDV(1)}
    \label{fig:1}
  \end{subfigure}
  \begin{subfigure}{0.5\textwidth}
    \centering
    \includegraphics[width=\linewidth]{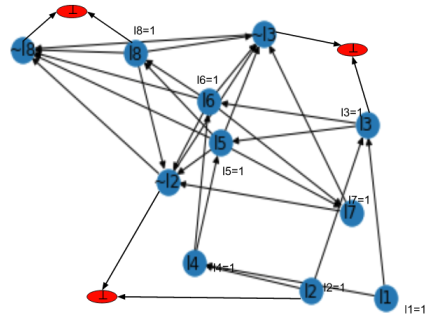}
    \caption{\scriptsize Implication graph for IDV(11)}
    \label{fig:2}
  \end{subfigure} 
  \begin{subfigure}{0.5\textwidth}\quad
    \centering
    \includegraphics[width=\linewidth]{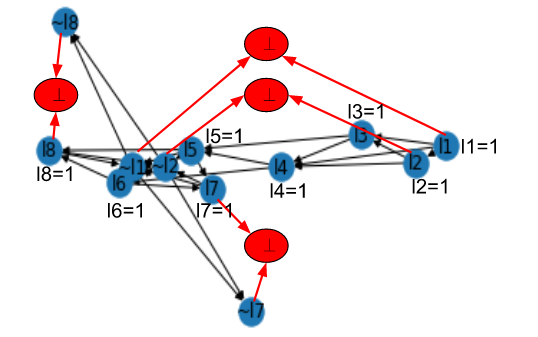}
    \caption{\scriptsize Implication graph for IDV(13)}
    \label{fig:3}
  \end{subfigure}
\caption{Conflict Analysis and Resolution Process}
  \label{Fig-model20}
\end{figure}

We denote conflicting bounds in red, as seen in Figure \ref{Fig-model20}. A conflicting safety bound occurs when a variable representing a lower or upper bound becomes true and false in the same domain as seen from the implication graph. The implication graph shows which variable is out-of-bound under the activation of disturbances. For example, the variable l2 represents the upper limit of the amount of feed A that should be in the reactor per hour. There is a violation of l2 from 30\% to 100\% with the activation of IDV(1), IDV(11) and IDV(13) resulting in a conflict. During the conflict analysis and resolution phase, the implication graph helps safety/security and engineering experts identify the main causes of violation of safety/security constraints and/or design goals to make an informed decision. 

From our conflict analysis and resolution, we found that the TE plant, although designed to operate properly under disturbances, goes into an unsafe state during the activation of certain disturbances. For example, the activation of IDV (1), IDV(11) and IDV(13) violates the bounded safety constraints of the plant during chemical production. The result of our analysis shows that disturbances 1, 11, and 13 should not be activated during chemical production of TE. This information and knowledge about the plant are necessary to redefine safety, security, or design goals to make the system more reliable, robust, and resilient to attacks.

\section{Conclusion and Future Work}
This paper proposes a methodology to address the concept of co-engineering of safety and security in ICS. In this work, we analyze the safety and security requirements of the Tennessee Eastman (TE) plant, where the safety requirements were represented as bounded constraints. The main goal of the paper is to identify, analyze, and resolve any conflicts/violations or vulnerabilities that occur during the safety and security analysis that an adversary can exploit. An attacker with proper knowledge of the system can conduct a well-orchestrated stealthy attack against the system by activating the right disturbances that can impact the system's behavior. Our study is relevant to ICS and other cyber-physical systems (CPS) fields with the aim of increasing the reliability, robustness, and resiliency of these systems against cyber attacks. In our result, we identified and analyzed the conflicts that arise from the safety and security boundary conditions in the TE plant during chemical production. Our analysis covers system integrity-related conflicts and/or violations, but not confidentiality-related and / or availability-related conflicts.

In the future, we will extend our proposed methodology to unbounded safety and security constraints, focusing mainly on confidentiality and/or availability-related conflicts. For example, security requirements ensure that the plant is shut down during an attack to prevent damage to the system. However, safety requirements ensure that the plant continues to run, provided that processes and/or process variables are within the desired limits.



\section*{Acknowledgment}
The authors would like to thank the National Science Foundation Computer and Information
Science and Engineering (CISE), award number 1846493 of the Secure and Trustworthy Cyberspace (SaTC)
program: Formal TOols foR SafEty aNd. Security of Industrial Control Systems (FORENSICS).

\bibliographystyle{IEEEtran}
\bibliography{IEEE_Italy}
\end{document}